\documentclass[12pt,english]{article}
\usepackage[T1]{fontenc}
\usepackage[latin9]{inputenc}
\usepackage{float}

\makeatletter

\providecommand{\tabularnewline}{\\}

\newcommand{\lyxaddress}[1]{
\par {\raggedright #1
\vspace{1.4em}
\noindent\par}
}

\makeatother

\usepackage{babel}
\begin{document}

\title{Neutrino Mass Spectrum for Co-Bimaximal Mixings from Quantum Gravity }

\author{Bipin Singh Koranga}

\maketitle

\lyxaddress{Department of Physics, Kirori Mal college (University of Delhi,) Delhi-110007,
India}
\begin{abstract}
We consider non-reormalizable interaction term as a perturbation of the
conventional neutrino mass matrix. Quantum gravitational (Planck scale
)effects lead to an effective $SU(2)_{L}\times U(1)$ invariant dimension-5
Lagrangian involving neutrino and Higgs fields,On  symmetry breaking, this
operator gives rise to correction to the neutrino masses and mixing. The
gravitational interaction $M_{X}=M_{pl}$ which gives rise to additional
terms in neutrino mass matrix. We also assume that, just above the electroweak
breaking scale, neutrino masses are nearly degenerate and their mixing
is Co-bimaximal mixing by assumming mixing angle $\theta_{13}\neq0=10^{o},\theta_{23}=\frac{\pi}{4},\,\,\, tan\theta_{12}^{2}=\frac{1-3sin\theta_{13}^{2}}{2}=34^{o}$
and Dirac phase$\delta=\pm\frac{\pi}{2}$.There additional term can be
considered to be perturbation of the GUT scale Co-bimaximal neutrino mass
matrix. The relation consider with solar and atmospheric neutrino oscillation
data predicted above GUT scale $m_{1}^{'}\simeq0.00001eV-0.00003eV,$ $m_{2}^{'}\simeq0.00008eV-0.00012,$and
$m_{3}^{'}\simeq0.000207eV-000320eV.$ 
\end{abstract}

\section{Introduction}

The phenemenon of neutrino oscillation, impressive advance have been made
to understand the phenomenology of neutrino oscillation through solar neutrino,atmospheric
neutrino,reactor neutrino and accelerator neutrino experiments. These experiments
enabling the determination of the Maki-Nakigawa-Sakata (MNS) lepton mixing
matrix.The main physical goal in future experiment are the determination
of the unknown parameter $\theta_{13}$ . In particular, the observation
of $\delta$ is quites interesting from the point of view that~$\delta$
related to the origin of the matter in the universe..CP violation is one
of the most important problem in particle physics {[}1,2,3,4,5{]}. One
of the most important parameter in neutrino physics is the magnitude of
mixing angle $\theta_{13}$ and CP phase $\delta$. The search for CP violation
is great interest to various aspects of neutrino physics. At present information
regarding $\theta_{13}$, we have an upper bound that is bases on the CHOOZ
experiment {[}6{]}, that is $\theta_{13}\leq13^{0}$ at 3 sigma. Solar
and atmospheric neutrino oscillation are assocated with the neutrino mass
square differences $\Delta_{21}=m_{2}^{2}-m_{1}^{2}$ and $\Delta_{31}=m_{3}^{2}-m_{1}^{2}$
respectively. In anology to the quark sector, $\theta_{12}$ is the $e-\mu$
mixing in the neutrino sector related to mass eigenstate as {[}7{]}

\begin{equation}
m_{1}^{2}=\frac{sin^{4}\theta_{12}}{cos2\theta_{12}}\Delta_{21}
\end{equation}

\begin{equation}
m_{2}^{2}=\frac{cos^{4}\theta_{12}}{cos2\theta_{12}}\Delta_{21}
\end{equation}

\begin{equation}
m_{3}^{2}=\frac{cos^{4}\theta_{12}}{cos2\theta_{12}}\Delta_{21}+\Delta_{32}
\end{equation}

The numerical dependence of two neutrino masses $m_{1}$ and $m_{2}$ on
the mixing angle $\theta_{12}.$In Ma{[}18{]},proposed a new mixing matrix
which is known as Co-bimaximal mixing by assumming mixing angle $\theta_{13}\neq0=10^{o},\theta_{23}=\frac{\pi}{4},\,\,\, tan\theta_{12}^{2}=\frac{1-3sin\theta_{13}^{2}}{2}$
and Dirac phase$\delta=\pm\frac{\pi}{2}$. Additional effects,which modify
the above predections,must exist so that the final predection are close
to the experimental determined valuesThe main purpose of the paper is to
study a possible relation between neutrino mixing and neutrino mass eigenstate
above the GUT scale. In Section 2, we outline the neutrino mixing parameter
above GUT scale. In Section 3, numerical results are given and section
4 is devoted to the conclusions.

\section{Neutrino Oscillation Parameter above GUT Scale }

The neutrino mass matrix is assumed to be generated by the see saw mechanism
{[}8, 9, 10{]}. We assume that the dominant part of neutrino mass matrix
arise due to GUT scale operators and the lead to Co-bimaximal mixing. The
effective gravitational interaction of neutrino with Higgs field can be
expressed as $SU(2)_{L}\times U(1)$ invariant dimension-5 operator {[}11{]},

\begin{equation}
L_{grav}=\frac{\lambda_{\alpha\beta}}{M_{pl}}(\psi_{A\alpha}\epsilon\psi_{C})C_{ab}^{-1}(\psi_{B\beta}\epsilon_{BD}\psi_{D})+h.c.
\end{equation}

Here and every where we use Greek indices $\alpha,\,\beta$ for the flavor
states and Latin indices i,j,k for the mass states. In the above equation
$\psi_{\alpha}=(\nu_{\alpha},l_{\alpha})$is the lepton doublet, $\phi=(\phi^{+},\phi^{o})$is
the Higgs doublet and $M_{pl}=1.2\times10^{19}GeV\,$is the Planck mass
$\lambda$ is a $3\times3$ matrix in a flavor space with each elements
$O(1)$. The Lorentz indices $a,\, b=1,2,3,4$ are contracted with the
charge conjugation matrix $C$ and the $SU(2)_{L}$ isospin indices $A,B,C,D=1,2$
are contracted with $\epsilon=i\sigma_{2},\,\,\sigma_{m}(m=1,2,3)$are
the Pauli matrices. After spontaneous electroweak symmetry breaking the
Lagrangian in Eq.(4.0) generated additional term of neutrino mass matrix

\begin{equation}
L_{mass}=\frac{v^{2}}{M_{pl}}\lambda_{\alpha\beta}\nu_{\alpha}C^{-1}\nu_{\beta},
\end{equation}

where $v=174GeV$ is the $VEV$ of electroweak symmetric breaking. We assume
that the gravitational interaction is''flavor blind'' that is $\lambda_{\alpha\beta}$
is independent of $\alpha,\,\beta\,\,$indices. Thus the Planck scale contribution
to the neutrino mass matrix is

\begin{equation}
\mu\lambda=\mu\left(\begin{array}{ccc}
1 & 1 & 1\\
1 & 1 & 1\\
1 & 1 & 1
\end{array}\right),
\end{equation}

where the scale $\mu$ is 

\begin{equation}
\mu=\frac{v^{2}}{M_{pl}}=2.5\times10^{-6}eV.
\end{equation}

We take Eq.(6.0) as perturbation to the main part of the neutrino mass
matrix, that is generated by GUT dynamics. To calculate the effects of
perturbation on neutrino observables. The calculation developed in an earlier
paper {[}13{]}. A natural assumption is that unperturbed ($0^{th}$ order
mass matrix) $M$~is given by

\begin{equation}
\mathbf{M}=U^{*}diag(M_{i})U^{\dagger},
\end{equation}

where, $U_{\alpha i}$ is the usual mixing matrix and $M_{i}$, the neutrino
masses is generated by Grand unified theory. Most of the parameter related
to neutrino oscillation are known, the major expectation is given by the
mixing elements $U_{e3}.$ We adopt the usual parametrization.

\begin{equation}
\frac{|U_{e2}|}{|U_{e1}|}=tan\theta_{12},
\end{equation}

\begin{equation}
\frac{|U_{\mu3}|}{|U_{\tau3}|}=tan\theta_{23},
\end{equation}

\begin{equation}
|U_{e3}|=sin\theta_{13}.
\end{equation}

In term of the above mixing angles, the mixing matrix is

\begin{equation}
U=diag(e^{if1},e^{if2},e^{if3})R(\theta_{23})\Delta R(\theta_{13})\Delta^{*}R(\theta_{12})diag(e^{ia1},e^{ia2},1).
\end{equation}

The matrix $\Delta=diag(e^{\frac{1\delta}{2}},1,e^{\frac{-i\delta}{2}}$)
contains the Dirac phase. This leads to CP violation in neutrino oscillation
$a1$ and $a2$ are the so called Majoring phase, which effects the neutrino
less double beta decay. $f1,$ $f2$ and $f3$ are usually absorbed as
a part of the definition of the charge lepton field. Planck scale effects
will add other contribution to the mass matrix that gives the new mixing
matrix can be written as {[}10{]}

\[
U^{'}=U(1+i\delta\theta),
\]

\[
\left(\begin{array}{ccc}
U_{e1} & U_{e2} & U_{e3}\\
U_{\mu1} & U_{\mu2} & U_{\mu3}\\
U_{\tau1} & U_{\tau2} & U_{\tau3}
\end{array}\right)
\]

\begin{equation}
+i\left(\begin{array}{ccc}
U_{e2}\delta\theta_{12}^{*}+U_{e3}\delta\theta_{23,}^{*} & U_{e1}\delta\theta_{12}+U_{e3}\delta\theta_{23}^{*}, & U_{e1}\delta\theta_{13}+U_{e3}\delta\theta_{23}^{*}\\
U_{\mu2}\delta\theta_{12}^{*}+U_{\mu3}\delta\theta_{23,}^{*} & U_{\mu1}\delta\theta_{12}+U_{\mu3}\delta\theta_{23}^{*}, & U_{\mu1}\delta\theta_{13}+U_{\mu3}\delta\theta_{23}^{*}\\
U_{\tau2}\delta\theta_{12}^{*}+U_{\tau3}\delta\theta_{23}^{*}, & U_{\tau1}\delta\theta_{12}+U_{\tau3}\delta\theta_{23}^{*}, & U_{\tau1}\delta\theta_{13}+U_{\tau3}\delta\theta_{23}^{*}
\end{array}\right).
\end{equation}

Where $\delta\theta$ is a hermit ion matrix that is first order in $\mu${[}12{]}.
The first order mass square difference $\Delta M_{ij}^{2}=M_{i}^{2}-M_{j}^{2},$get
modified {[}10{]} as

\begin{equation}
\Delta M_{ij}^{'^{2}}=\Delta M_{ij}^{2}+2(M_{i}Re(m_{ii})-M_{j}Re(m_{jj}),
\end{equation}

where

\[
m=\mu U^{t}\lambda U,
\]

\[
\mu=\frac{v^{2}}{M_{pl}}=2.5\times10^{-6}eV.
\]

The change in the elements of the mixing matrix, which we parametrized
by $\delta\theta${[}10{]}, is given by

\begin{equation}
\delta\theta_{ij}=\frac{iRe(m_{jj})(M_{i}+M_{j})-Im(m_{jj})(M_{i}-M_{j})}{\Delta M_{ij}^{'^{2}}}.
\end{equation}

The above equation determine only the off diagonal elenumerments of matrix
$\delta\theta_{ij}$. The diagonal element of $\delta\theta_{ij}$ can
be set to zero by phase invariance. Using Eq(13), we can calculate neutrino
mixing angle due to Planck scale effects,

\begin{equation}
\frac{|U_{e2}^{'}|}{|U_{e1}^{'}|}=tan\theta_{12}^{'},
\end{equation}

\begin{equation}
\frac{|U_{\mu3}^{'}|}{|U_{\tau3}^{'}|}=tan\theta_{23}^{'},
\end{equation}

\begin{equation}
|U_{e3}^{'}|=sin\theta_{13}^{'}
\end{equation}

For degenerate neutrinos, $M_{3}-M_{1}\cong M_{3}-M_{2}\gg M_{2}-M_{1},$
because $\Delta_{31}\cong\Delta_{32}\gg\Delta_{21}.$ Thus, from the above
set of equations, we see that $U_{e1}^{'}$ and $U_{e2}^{'}$ are much
larger than $U_{e3}^{'},\,\, U_{\mu3}^{'}$ and $U_{\tau3}^{'}$. Hence
we can expect much larger change in $\Delta_{21}${[}14{]} and $\theta_{12}$
compared to $\theta_{13}$ and $\theta_{23}\,[13].$ As one can see from
the above expression of mixing angle due to Planck scale effects, depends
on new contribution of mixing matrix $U^{'}=U(1+i\delta\theta).$ The above
statements are not dependent on the exact from of the matrix $\lambda$
given in eq(6). They hold true for any form of $\lambda$, as long as all
its elements are of order 1. New contribution of mixing following value
of neutrino mass above GUT scale using Eq.(09) to Eq.(11)

\begin{equation}
m_{1}^{'2}=\frac{sin^{4}\theta_{12}^{'}}{cos2\theta_{12}^{'}}\Delta_{21}^{'}
\end{equation}

\begin{equation}
m_{2}^{'2}=\frac{cos^{4}\theta_{12}^{'}}{cos2\theta_{12}^{'}}\Delta_{21}^{'}
\end{equation}

\begin{equation}
m_{3}^{'2}=\frac{cos^{4}\theta_{12}^{'}}{cos2\theta_{12}^{'}}\Delta_{21}^{'}+\Delta_{32}^{'}
\end{equation}

\section{Numerical Results}

We assume that, just above the electroweak breaking scale, the neutrino
masses are nearly degenerate and the mixing are Co-bimaximal by assumming
mixing angle $\theta_{13}\neq0=10^{o},\theta_{23}=\frac{\pi}{4},\,\,\, tan\theta_{12}^{2}=\frac{1-3sin\theta_{13}^{2}}{2}=34^{o}$
and Dirac phase$\delta=\pm\frac{\pi}{2}$. Taking the common degenerate
neutrino mass to be 2 eV, which is the upper limit coming from tritium
beta decay {[}15{]}. We compute the modified mixing angles using Eq. (19)
to Eq.(21). We have taken $\Delta_{31}=0.002eV^{2}[16]$ and $\Delta_{21}=0.00008eV^{2}${[}17{]}.
For simplicity we have set the charge lepton phases $f_{1}=f_{2}=f_{3}=0.$
In table-1, we list the modified neutrino mixing angles and neutrino masses
for some sample value of $a1$ and $a2$ . Due to Planck scale effects,
only $\theta_{12}$ have reasonable deviation and $\theta_{23},\,\theta_{13}$
deviation is very small less then $0.3^{o}${[}13{]}. From table-1, we
list the neutrino masses for Co-bimaximal mixing pattern $\theta_{13}\neq0=10^{o},\theta_{23}=\frac{\pi}{4},\,\,\, tan\theta_{12}^{2}=\frac{1-3sin\theta_{13}^{2}}{2}=34^{o}$
and Dirac phase$\delta=\pm\frac{\pi}{2}$. In table-1 and table-2, we have
listed modified neutrino masses Vs reasonable range of Majarona phases.

\begin{table}[H]
\begin{tabular}{|c|c|c|c|c|}
\hline 
$a1$ & $a2$ & $m_{1}^{'}$in eV & $m_{2}^{'}$in eV & $m_{3}^{'}$in eV\tabularnewline
\hline 
\hline 
$0^{o}$ & $0^{o}$ & 0.00003 & 0.00011 & 0.00317\tabularnewline
\hline 
$0^{o}$ & $45^{o}$ & 0.00003 & 0.00010 & 0.00302\tabularnewline
\hline 
$0^{o}$ & $90^{o}$ & 0.00001 & 0.00008 & 0.00267\tabularnewline
\hline 
$0^{o}$ & $135^{o}$ & 0.00002 & 0.00009 & 0.00277\tabularnewline
\hline 
$0^{o}$ & $180^{o}$ & 0.00003 & 0.00011 & 0.00317\tabularnewline
\hline 
$45^{o}$ & $0^{o}$ & 0.00003 & 0.00012 & 0.00303\tabularnewline
\hline 
$45^{o}$ & $45^{o}$ & 0.00002 & 0.00011 & 0.00290\tabularnewline
\hline 
$45^{o}$ & $90^{o}$ & 0.00001 & 0.00009 & 0.00262\tabularnewline
\hline 
$45^{o}$ & $135^{o}$ & 0.00002 & 0.00010 & 0.00271\tabularnewline
\hline 
$45^{o}$ & $180^{o}$ & 0.00003 & 0.00012 & 0.00303\tabularnewline
\hline 
$90^{o}$ & $0^{o}$ & 0.00002 & 0.00011 & 0.00270\tabularnewline
\hline 
$90^{o}$ & $45^{o}$ & 0.00001 & 0.00010 & 0.00263\tabularnewline
\hline 
$90^{o}$ & $90^{o}$ & 0.00001 & 0.00009 & 0.00246\tabularnewline
\hline 
$90^{o}$ & $135^{o}$ & 0.00001 & 0.00009 & 0.00251\tabularnewline
\hline 
$90^{o}$ & $180^{o}$ & 0.00002 & 0.00011 & 0.00270\tabularnewline
\hline 
$135^{o}$ & $0^{o}$ & 0.00002 & 0.00010 & 0.00276\tabularnewline
\hline 
$135^{o}$ & $45^{o}$ & 0.00001 & 0.00009 & 0.00268\tabularnewline
\hline 
$135^{o}$ & $90^{o}$ & 0.00001 & 0.00008 & 0.00248\tabularnewline
\hline 
$135^{o}$ & $135^{o}$ & 0.00001 & 0.00009 & 0.00254\tabularnewline
\hline 
$135^{o}$ & $180^{o}$ & 0.00002 & 0.00010 & 0.00276\tabularnewline
\hline 
$180^{o}$ & $0^{o}$ & 0.00003 & 0.00011 & 0.00317\tabularnewline
\hline 
$180^{o}$ & $45^{o}$ & 0.00003 & 0.00010 & 0.00302\tabularnewline
\hline 
$180^{o}$ & $90^{o}$ & 0.00001 & 0.00008 & 0.00267\tabularnewline
\hline 
$180^{o}$ & $135^{o}$ & 0.00002 & 0.00009 & 0.00277\tabularnewline
\hline 
$180^{o}$ & $180^{o}$ & 0.00003 & 0.00011 & 0.00317\tabularnewline
\hline 
\end{tabular}

\caption{The modified neutrino mass square difference term for various value of
phase. Input value are $\Delta_{31}=2.0\times10^{-3}eV^{2}$, $\Delta_{21}=8.0\times10^{-5}eV^{2}$and
Co-bimaximal mixing by assumming mixing angle $\theta_{13}\neq0=10^{o},\theta_{23}=\frac{\pi}{4},\,\,\, tan\theta_{12}^{2}=\frac{1-3sin\theta_{13}^{2}}{2},\theta_{12}=34^{o}$
and Dirac phase$\delta=+\frac{\pi}{2}$.}
\end{table}

\begin{table}[H]
\begin{tabular}{|c|c|c|c|c|}
\hline 
$a1$ & $a2$ & $m_{1}^{'}$in eV & $m_{2}^{'}$in eV & $m_{3}^{'}$in eV\tabularnewline
\hline 
\hline 
$0^{o}$ & $0^{o}$ & 0.00003 & 0.00011 & 0.00317\tabularnewline
\hline 
$0^{o}$ & $45^{o}$ & 0.00002 & 0.00009 & 0.00276\tabularnewline
\hline 
$0^{o}$ & $90^{o}$ & 0.00001 & 0.00008 & 0.00267\tabularnewline
\hline 
$0^{o}$ & $135^{o}$ & 0.00003 & 0.00010 & 0.00303\tabularnewline
\hline 
$0^{o}$ & $180^{o}$ & 0.00003 & 0.00011 & 0.00317\tabularnewline
\hline 
$45^{o}$ & $0^{o}$ & 0.00002 & 0.00010 & 0.00276\tabularnewline
\hline 
$45^{o}$ & $45^{o}$ & 0.00001 & 0.00009 & 0.00254\tabularnewline
\hline 
$45^{o}$ & $90^{o}$ & 0.00001 & 0.00008 & 0.00249\tabularnewline
\hline 
$45^{o}$ & $135^{o}$ & 0.00001 & 0.00009 & 0.00268\tabularnewline
\hline 
$45^{o}$ & $180^{o}$ & 0.00002 & 0.00010 & 0.00276\tabularnewline
\hline 
$90^{o}$ & $0^{o}$ & 0.00002 & 0.00011 & 0.00270\tabularnewline
\hline 
$90^{o}$ & $45^{o}$ & 0.00001 & 0.00010 & 0.00252\tabularnewline
\hline 
$90^{o}$ & $90^{o}$ & 0.00001 & 0.00009 & 0.00247\tabularnewline
\hline 
$90^{o}$ & $135^{o}$ & 0.00001 & 0.00010 & 0.00264\tabularnewline
\hline 
$90^{o}$ & $180^{o}$ & 0.00002 & 0.00011 & 0.00270\tabularnewline
\hline 
$135^{o}$ & $0^{o}$ & 0.00003 & 0.00012 & 0.00303\tabularnewline
\hline 
$135^{o}$ & $45^{o}$ & 0.00002 & 0.00010 & 0.00270\tabularnewline
\hline 
$135^{o}$ & $90^{o}$ & 0.00001 & 0.00009 & 0.00262\tabularnewline
\hline 
$135^{o}$ & $135^{o}$ & 0.00003 & 0.00011 & 0.00309\tabularnewline
\hline 
$135^{o}$ & $180^{o}$ & 0.00004 & 0.00012 & 0.00320\tabularnewline
\hline 
$180^{o}$ & $0^{o}$ & 0.00003 & 0.00011 & 0.00317\tabularnewline
\hline 
$180^{o}$ & $45^{o}$ & 0.00002 & 0.00009 & 0.00276\tabularnewline
\hline 
$180^{o}$ & $90^{o}$ & 0.00001 & 0.00008 & 0.00267\tabularnewline
\hline 
$180^{o}$ & $135^{o}$ & 0.00003 & 0.00010 & 0.00303\tabularnewline
\hline 
$180^{o}$ & $180^{o}$ & 0.00003 & 0.00011 & 0.00317\tabularnewline
\hline 
\end{tabular}

\caption{The modified neutrino mass square difference term for various value of
phase. Input value are $\Delta_{31}=2.0\times10^{-3}eV^{2}$, $\Delta_{21}=8.0\times10^{-5}eV^{2}$and
Co-bimaximal mixing by assumming mixing angle $\theta_{13}\neq0=10^{o},\theta_{23}=\frac{\pi}{4},\,\,\, tan\theta_{12}^{2}=\frac{1-3sin\theta_{13}^{2}}{2},\theta_{12}=34^{o}$
and Dirac phase$\delta=-\frac{\pi}{2}$.}
\end{table}

\section{Conclusions}

We assume that the main part of neutrino masses and mixing arise from GUT
scale operator. We considered these to be $0^{th}$ order quantities. The
gravitational interaction of lepton field with SM Higgs field give rise
to a $SU(2)_{L}\times U(1)$ invariant dimension-5 effective Lagrangian
give originally by Weinberg {[}11{]}. On electroweak symmetry breaking
this operators leads to additional mass terms. We considered these to be
perturbation of GUT scale mass terms. This model predicts the modified
neutrino masses $m_{1}^{'}=0.00001eV-0.00003eV,$ $m_{2}^{'}=0.00008eV-0.00012eV$
and $m_{3}^{'}=0.000207eV-0.000320,$$,$which is corresponds to Planck
scale $M_{pl}\approx2.0\times10^{19}GeV.$ In this paper, we studied how
physics from planck scale effects the neutrino masses. We compute the modified
neutrino mass eigenvalues due to the additional mass terms for the case
of Co-bimaximal mixing. The change in $\Delta_{31},$due to these Planck
scale correction are negligible. But the change in $\Delta_{21}^{'}$ is
enough that final value falls within the expermentally accepted region
{[}19{]}. This occurs,of course for degenerate neutrino mass with a common
mass of about 2eV.

\end{document}